\documentclass[pre,aps,twocolumn,amsmath,amssymb,showpacs]{revtex4-1}
\usepackage{graphicx}
\usepackage{comment}

\begin{document}
\title{Collapse transition in polymer models with multiple monomers per site and multiple bonds per edge}
\author{Nathann T. Rodrigues}
\email{nathan.rodrigues@ufv.br}
\author{Tiago J. Oliveira}
\email{tiago@ufv.br}
\affiliation{Departamento de F\'isica, Universidade Federal de Vi\c cosa, 36570-900, Vi\c cosa, MG, Brazil}
\date{\today}

\begin{abstract}
We present results from extensive Monte Carlo simulations of polymer models where each lattice site can be visited by up to $K$ monomers and no restriction is imposed on the number of bonds on each lattice edge. These \textit{multiple monomer per site} (MMS) models are investigated on the square and cubic lattices, for $K=2$ and $K=3$, by associating Boltzmann weights $\omega_0=1$, $\omega_1=e^{\beta_1}$ and $\omega_2=e^{\beta_2}$ to sites visited by 1, 2 and 3 monomers, respectively. Two versions of the MMS models are considered for which immediate reversals of the walks are allowed (RA) or forbidden (RF). In contrast to previous simulations of these models, we find the same thermodynamic behavior for both RA and RF versions. In three-dimensions, the phase diagrams - in space $\beta_2 \times \beta_1$ - are featured by coil and globule phases separated by a line of $\Theta$ points, as thoroughly demonstrated by the metric $\nu_t$, crossover $\phi_t$ and entropic $\gamma_t$ exponents. The existence of the $\Theta$-lines is also confirmed by the second virial coefficient. This shows that no discontinuous collapse transition exists in these models, in contrast to previous claims based on a weak bimodality observed in some distributions, which indeed exists in a narrow region very close to the $\Theta$-line when $\beta_1 < 0$. Interestingly, in two-dimensions, only a crossover is found between the coil and globule phases.
\end{abstract}


\maketitle

\section{Introduction}
\label{intro}

Lattice (random) walks have long been used as simplified models for linear polymers in solution. Usually, the monomers are represented by a sequence of nearest-neighbor (NN) lattice sites which are connected by bonds placed on the lattice edges \cite{Carlo,DeGennes2,Cloizeaux}. A key ingredient in such modeling is the inclusion of self-avoidance, to represent the excluded volume present in real systems, which turns the walks a non-trivial problem and place them in a central stage into Statistical Mechanics. The simplest way to consider this exclusion is imposing that each lattice \textit{site} can be occupied by at most one \textit{monomer}, leading to the so-called \textit{self-avoiding walks} (SAWs). On the other hand, when the exclusion is imposed on the lattice \textit{edges}, so that each one can be occupied by at most one \textit{bond}, one has \textit{bond-avoiding walks} (BAWs), aka trails.

Beyond the excluded volume, the inclusion of self-attraction in the walks is also worthy to model collapsing polymers \cite{Flory1,Flory2}. In SAWs, this is usually done by introducing an attractive interaction between non-bonded NN monomers - yielding the celebrated interacting SAW (ISAW) model \cite{Carlo,Flory2,DeGennes2} - which indeed undergoes a collapse transition at the so-called $\Theta$-point. More specifically, for high temperatures $T \gg T_{\Theta}$ (or good solvents) the effect of the excluded volume dominates and the polymer chains are swollen, similarly to ordinary SAWs. In opposition, for $T\ll T_{\Theta}$ (or poor solvents) they are collapsed in globular conformations, due to the dominance of self-attraction. At $T=T_{\Theta}$ (or in a ``$\Theta$-solvent'') there is a balance (on average) between exclusion and attraction, giving rise to ideal chains - the $\Theta$ polymers. From the $O(n)$ field theory (in the limit $n \rightarrow 0$) the $\Theta$-point is recognized as a tricritical point \cite{DeGennes1,DeGennes2}. In three dimensions (3D), which is the upper critical dimension of this transition, the $\Theta$ exponents are expected to assume mean-field values \cite{Flory2,DeGennes2} except by logarithmic corrections \cite{Duplantier1,Duplantier2,Grassberger3d}. In 2D, notwithstanding, the situation is more complex and the generic critical behavior at the $\Theta$-point have been subject of a long debate (see e.g., Refs. \cite{vernier15,nahum16} for detailed discussions), but several numerical evidences \cite{DupSaleur2,Prellber94,Caracciolo11,Nathann14} suggest that the robust critical exponents are those derived by Duplantier and Saleur (DS) \cite{DupSaleur}, provided that the walks do not cross themselves \cite{nahum16}.

Collapse (coil-globule) transitions have also been found in BAW models where attractive interactions are \textit{on-site}, i.e., associated to monomers at multiply visited sites rather than to NN ones. This is the case, for example, in the interacting self-avoiding trail (ISAT) model \cite{massih75} - where each lattice edge can have at most one bond, while the number of monomers per site is limited only by the lattice coordination. When crossings are forbidden in the ISAT model on the square lattice, the so-called vertex interacting SAW (VISAW) model \cite{bn89} is obtained. Bl\"ote and Nienhuis (BN) \cite{bn89} solved this model and found a tricritical point different from the DS one, and it was suggested in  \cite{nienhuis92} that it could be the generic $\Theta$-point. However, recent evidences against this exists coming from numerical \cite{Foster12,bedini13} and field theory \cite{vernier15} works. Controversies exist also on the ISAT collapse transition \cite{onno84,lyklema85,lim88,chang92,Foster09,Prellber95,Grassberger96}, which have motivated several recent works on this model and generalizations of it \cite{bedini12,bedini13b,bedini13c,nahum13,TJ16,Pretti16,WTJP17}. Noteworthy among these works is the field theory by Nahum \textit{et al.} \cite{nahum13} showing that the ISAT collapse transition in 2D is multicritical with infinite order.

Another interesting class of models for polymer collapse with multiple monomers per site (MMS) and on-site interactions only was introduced one decade ago by Krawczyk \textit{et al.} \cite{Krawczyk}, inspired in the Domb-Joyce model. In such MMS models, self-avoidance is introduced by imposing that each lattice site can be occupied by at most $K$ monomers and, in contrast to BAWs, \textit{no} restriction exists on the number of bonds on each lattice edge (beyond that naturally imposed by $K$). Actually, apart from the on-site interactions, this is exactly the so-called \textit{$K$-tolerant walks}, first defined and discussed by Malakis \cite{Malakis} and further investigated in several 80's papers \cite{Turban83,onno84,Guttmann84,Dekeyser,Rieger88} focusing on the comparison of these walks with SAWs and BAWs. In the MMS model, the maximal number of possible bonds on each lattice edge depends also on whether immediate reversals of the walk are allowed (RA) or forbidden (RF). Krawczyk \textit{et al.} \cite{Krawczyk} studied both RA and RF models with $K=3$ on the square and simple cubic lattices, via Monte Carlo simulations, by assigning Boltzmann weights $\omega_0 = 1$, $\omega_1 = e^{\beta_1}$ and $\omega_2 = e^{\beta_2}$ to sites visited by one, two and three monomers, respectively. Even though very few data were explicitly shown in \cite{Krawczyk}, it was claimed there that the existence of the collapse transition in these models depends sensitively on their details and on system dimension. In short, for the RF model in 3D a rich phase diagram [in space ($\beta_1,\beta_2$)] was found, with coil and globule phases separated by continuous and discontinuous transition lines ``which possibly meet at a multicritical point, possibly located in the region of attractive interactions'' \cite{Krawczyk}. Regarding the universality class of the continuous transition line, Krawczyk \textit{et al.} only said that ``It may be the case that it is of the same type as ISAW collapse in three dimensions'', but no evidence of this was presented. Interestingly, for the model RA in 2D no evidence of a phase transition was found, but only a smooth crossover. Furthermore, it was concluded in \cite{Krawczyk} that models RA in 3D and RF in 2D display similar thermodynamic behavior, with a collapse transition existing in the region of $\beta_1 < 0$ (whose order was not explicitly stated), but with inconclusive results for the rest of the parameter space.

In contrast with these numerical results, the same thermodynamic behavior has been found for RA and RF models in exact solutions of them on hierarchical (Bethe and Husimi) lattices \cite{Pablo,Tiago}. In such solutions, coil and globule phases are always separated by lines of continuous transitions, being a tricritical line in the region of $\beta_2 < 0$ (with $\beta_1>0$) and a line of critical-end-points (CEP) for $\beta_1<0$ (with $\beta_2>0$), both meeting at a multicritical point \cite{Tiago}. As an aside, note that the region $\beta_1 > 0$ and $\beta_2< 0$ is somewhat related to the field theory by des Cloizeaux and Duplantier \cite{desCloiz75,Duplantier3} considering attractive (repulsive) two- (three-) body interactions. Since the hierarchical lattices are mean-field (MF) approximations for the models on regular ones, they yield classical critical exponents and, so, they do not provide the universality class of the transitions. Moreover, it is not possible to analyze the effect of dimensionality on the behavior of the models, because these lattices have infinite dimension. Anyhow, the results in \cite{Pablo,Tiago} strongly suggest that the collapse transitions in MMS models: \textit{i}) are \textit{not} dependent on their details; and \textit{ii}) are always continuous.

In order to resolve these controversies on these MMS models, we investigate them here through extensive Monte Carlo simulations (for walks' lengths $10$  times larger than those studied in \cite{Krawczyk}) on the square and simple cubic lattices. From the analysis of scaling exponents, as well as of the second virial coefficient we find strong evidences that RA and RF models have always similar phase diagrams. In 3D, they are featured by a $\Theta$-line separating coil and globule phases in the entire region of parameters analyzed. In 2D, on the other hand, only a crossover is observed (for both models).

The rest of this work is organized as follows. In Sec. \ref{defmod} we define the model, give some details on the simulation method and define the main quantities calculated. The thermodynamic behavior of the models on the cubic and square lattices are presented in Secs. \ref{results3D} and \ref{results2D}, respectively. In Sec. \ref{conclusions} our final discussions and conclusions are summarized.

\section{Models and quantities of interest}
\label{defmod}

We investigate the multiple monomer per site (MMS) model, aka $K$-tolerant walks, where each lattice site can be occupied by up to $K$ monomers. Following Krawczyk \textit{et al.} \cite{Krawczyk}, on-site interactions are introduced by assigning energies $\varepsilon_{i-1}$ to sites with $i$ monomers, with $\varepsilon_{i-1}=0$ for $i \leq 1$. Thereby, walks composed only by sites with a single monomer (i.e., ordinary SAWs) - have a null energy here. Hence, the energy of a given configuration $S$, with $N$ steps, reads 
\begin{equation}
E_N(S) = - \sum_{i=2}^K M_{i}(S) \varepsilon_{i-1},
\end{equation}
where $M_{i}(S)$ is the number of sites occupied by $i$ monomers in the walk $S$. Here, we will restrict ourselves to cases with $K \leq 3$, so that $E_N(S) = - M_{2}(S) \varepsilon_{1} - M_{3}(S) \varepsilon_{2}$. Two versions of the MMS model will be considered, according to the possibility of \textit{immediate} reversals of the walks:
\begin{itemize}
 \item for reversals allowed (RA), a walk can visit a site $j$, then one of its nearest-neighbors (NN) $k$ and immediately return to $j$;
 \item for reversals forbidden (RF), a sequence of the type $j-k-j$ cannot occur in the walks. 
\end{itemize}
Note that each lattice edge can have up to $2K-1$ bonds in RA model and $K$ bonds in RF one and, so, the ensemble of walks (i.e., the number of allowed configurations for a given $K$ and $N$) is larger in RA case. Following the notation in \cite{Krawczyk,Tiago}, we define the parameters $\beta_1\equiv\varepsilon_1/k_B T$ and $\beta_2\equiv\varepsilon_2/k_B T$, where $k_B$ is the Boltzmann's constant and $T$ is the temperature. So, the canonical partition function of a system with walks of $N$ steps is $Z_N = \sum_S e^{-E(S)/k_B T} = \sum_S e^{M_2(S) \beta_1 + M_3(S) \beta_2}$. 

To obtain an estimative of $Z_N$ and other relevant quantities one uses the pruned enriched Rosenbluth method (PERM) \cite{Grassberger1}, which is a powerful Monte Carlo method to sample long polymers chains, as well as a sort of other systems \cite{Grassberger2}. Indeed, we analyze here walks with up to $N=10000$ steps, with up to $10^8$ (and at least $10^6$) samples in the statistics for each set of parameters ($\beta_1,\beta_2$). For detailed descriptions of the method see, e.g., Refs. \cite{Rensburg,Grassberger2}. Here, we recall only that PERM is based on the classical Rosenbluth-Rosenbluth method \cite{Rosenbluth}, where walks are grown in a biased way by trying to insert new monomers at \textit{available} NN sites at their ends. Note that while available means empty sites in SAWs, in the MMS models it means sites occupied by $i < K$ monomers. To correct the statistics, a weight $W_N(S)$ is associated with each generated walk. A simple way to do this is as follows. If $l_n$ is the number of available NN sites at the step $n-1$, the $j^{th}$ of these sites is chosen with probability $p_j = 1/l_n$, and a ``local'' weight $w_n^{(j)} = l_{n} e^{-E_j/k_B T}$ is associated to step $n$. Thence, the Rosenbluth weight of a configuration $S$ with $N$ steps is given by $W_{N}(S) = \prod_{n=1}^{N} w_{n}^{(j_n)}$. Note that when a walk becomes trapped, so that $l_n=0$, it will have $W_N=0$. Thereby, if $L$ and $I_N$ denotes respectively the numbers of walks started and walks successfully generated (with $N$ monomers), the partition function of the system will be given by $Z_N \simeq \sum_{i=1}^{L} W(S_i)/I_N$ and, so, the expected value of an observable $A_N$ is obtained from $\left\langle A_N \right\rangle = \sum_{i=1}^{L} A(S_i) W(S_i)/\sum_{i=1}^{L} W(S_i)$. In the PERM algorithm, everything proceeds as above, but at each stage of the growth of a walk it can be either duplicated (whenever its weight becomes larger than a parameter $T_n$) or pruned (if its weight is smaller than a parameter $t_n$). We set $T_n/t_n = 10$ \cite{Grassberger1} in our simulations.

By keeping $\beta_j$ fixed (with $j=1$ or $2$) and varying $\beta_i$ (with $i=2$ or 1, respectively), the average end-to-end distance, $R_N$, (and similarly the radius of gyration) of a chain with $N$ steps is expected to scale near the criticality as \cite{DeGennes1,DeGennes2}
\begin{equation}
 \left\langle R_N^2 \right\rangle \sim N^{2 \nu_t} f\left( \tau_i N^{\phi_t} \right),
\label{eqScaling}
\end{equation}
where $\nu_t$ and $\phi_t$ are the tricritical metric and crossover exponents, respectively, $\tau_i \equiv |\beta_i - \beta_{i,\Theta}|$, and $f(x)$ is a scaling function, expected to behave as \cite{meirovitch,meirovitch2,Tesi}
\begin{equation}
f(x) \sim \left \{
\begin{array}{ll}
x^{(2 \nu_{SAW} -2\nu_t)/\phi_t}, & \text{if} \quad x \rightarrow \infty, \\
const., & \text{if} \quad x=0, \\
|x|^{(2/d-2\nu_t)/\phi_t}, & \text{if} \quad x \rightarrow -\infty.
\end{array}
\right.
\label{eqScaling2}
\end{equation}
Therefore, for fixed $(\beta_1,\beta_2)$, one must have $\left\langle R_N^2 \right\rangle \sim N^{2 \nu}$, with exponents $\nu=\nu_{SAW}$ in the coil phase, $\nu=\nu_t$ at the $\Theta$-point and $\nu=1/d$ in the globule phase. Hence, by calculating metric exponents from $\ln \left\langle R_N^2 \right\rangle \times \ln N$ scale for different $\beta_i$ and lengths $N$, they are expected to have the same value at the $\Theta$-point, so that curves of $\nu(\beta_i) \times \beta_i$ for different $N$ should intersect each other at $(\beta_{i,\Theta},\nu_t)$. Similarly, curves of rescaled end-to-end distance $\left\langle R_N^2 \right\rangle/N^{2 \nu_t}$ versus $\beta_i$, for different $N$'s, are expected to intersect at a single point - the $\Theta$-point - provided that the correct exponent $\nu_t$ is used. In 3D, the tricritical exponents are believed to assume the mean-field values $\nu_{t}=\phi_t=1/2$ \cite{Flory2,DeGennes2}. In 2D, the Duplantier-Saleur (DS) exponents have the values $\nu_{t}^{(DS)}=4/7$ and $\phi_t^{(DS)}=3/7$ \cite{DupSaleur}, while the metric exponent for the Bl\"ote-Nienhus (BN) tricritical point is $\nu_t^{(BN)}=12/23$ \cite{bn89}.

From Eq. \ref{eqScaling}, the derivative $R'_N \equiv \partial \ln \left\langle R_N^2 \right\rangle/\partial \tau_i$ is expected to scale, at the $\Theta$-point, as
\begin{equation}
 R'_N \sim N^{\phi_t},
\label{eqScalingphi}
\end{equation}
so that the exponent $\phi_t$ can be estimated from plots of $\ln R'_N \times \ln N$ (at the $\Theta$-point). We calculate $R'_N$ numerically here from $R'_N = [\ln \left\langle R_N^2 \right\rangle(\beta_{i,\Theta}+\Delta_i)-\ln \left\langle R_N^2 \right\rangle(\beta_{i,\Theta}-\Delta_i)] / 2 \Delta_i$ and robust values were found for different $\Delta_i (\ll 1)$ and for $i=1$ or $2$.

For $\beta_i \ge \beta_{i,\Theta}$, the partition function of the system is expected to scale as
\begin{equation}
Z_N \sim \mu^N N^{\gamma-1},
\end{equation}
where $\mu$ is the connectivity coefficient and $\gamma$ is the entropic exponent, which can be calculated from the expression $2 Z_{2 N}/( Z_N \mu^N ) = 2^{\gamma}$. At the $\Theta$-point, $\gamma_t=1$ in 3D, while in 2D it is known that $\gamma_t^{(DS)}=8/7$ \cite{DupSaleur} and $\gamma_t^{(BN)}=53/46$ \cite{bn89}.

Since at the $\Theta$-point the polymer chains should display an ideal behavior in 3D, in an expansion of the osmotic pressure $\Pi$ for low monomer concentration $\rho$, namely, $\Pi/R T = \rho/N + A_2 \rho^2 + O(\rho^3)$, the second virial coefficient $A_2$ shall vanish at the $\Theta$-point (when $N \rightarrow \infty$). Indeed, $A_2(\beta_1,\beta_2,N \rightarrow \infty)=0$ is the simplest and more physical definition of the $\Theta$-point ($\beta_{1,\Theta},\beta_{2,\Theta}$) \cite{Cloizeaux}. For finite walks, one has $A_2=0$ at the Boyle temperature $T^*(N)$ [a ``Boyle parameter'' $\beta_i^*(N)$ in our case] and then \cite{Janssens,Grassberger3d}
\begin{equation}
\beta_i^*(N) - \beta_{i,\Theta} \simeq b N^{-1/2}.
\label{eqScalingA2}
\end{equation}
The second virial coefficient can be calculated as 
\begin{equation}
A_2 = -\frac{Z_N^{C}}{2 N^2 Z_N^2},
\end{equation}
where $Z_N^C$ is the partition function relative to two chains \cite{Cloizeaux}. To calculate it, we follow the same procedure done in Refs. \cite{Grassberger3d,Janssens}. Namely, a pair of walks (say 1 and 2, both for the same parameters $\beta_1$ and $\beta_2$ and length $N$) are independently generated and then placed together on the lattice with origins at positions $O_1$ and $O_2$. While $O_1$ is kept fixed, $O_2$ is variated in the neighborhood of chain $1$, assuming all positions that yield a superposition of the two chains. For each of these positions, we determine the points where walks 1 and 2 cross each other. Note that the superposition constraint implies that they shall cross at least in one point. If one of these crossing points has four or more monomers, we will say that the chains (for that $O_2$) are ``overlapped'' in MMS models with $K=3$ and then the counter of ``overlappings'' $C_O$ is appropriately increased. Otherwise, we determine how many of the crossing sites has two and three monomers and increase their counters ($C_2$ and $C_3$, respectively) according. After generating a very large number (at least $10^6$) of pairs of walks and varying $O_2$ for each one of them, accurate estimates of $\left\langle C_O \right\rangle$, $\left\langle C_2 \right\rangle$ and $\left\langle C_3 \right\rangle$ are obtained, from which $A_2$ is calculated following Ref. \cite{Janssens}. Note that when a crossing site has two monomers their interaction is simply $\beta_1$, but if it has three monomers the net two-chain interaction will be given by $\beta_2-\beta_1$.

\section{Results for the cubic lattice}
\label{results3D} 

In this section, results from simulations of the RA and RF models on the simple cubic lattice are presented.

\subsection{$K=2$}

\begin{figure}[t]
\includegraphics[width=8.cm]{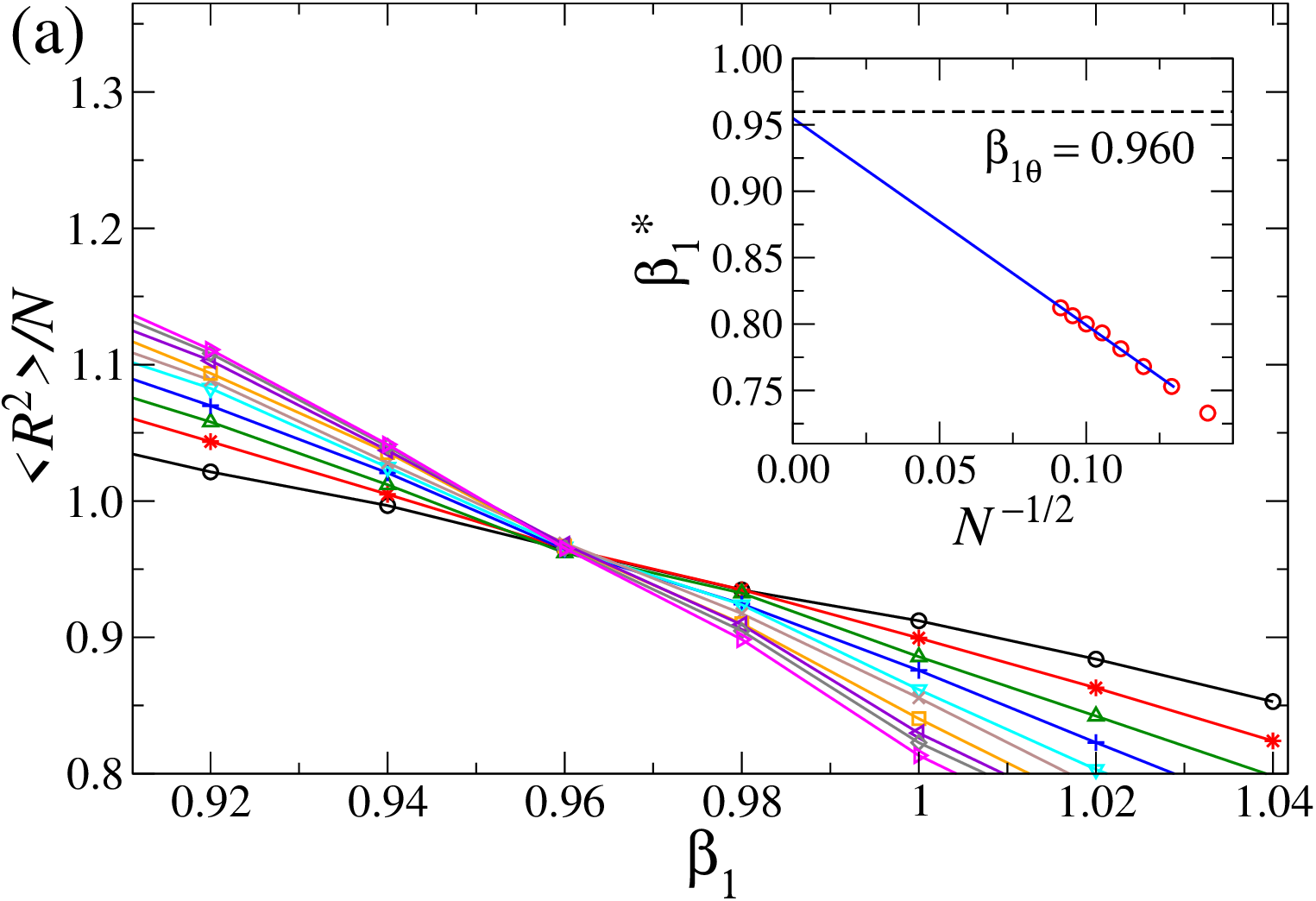}
\includegraphics[width=8.cm]{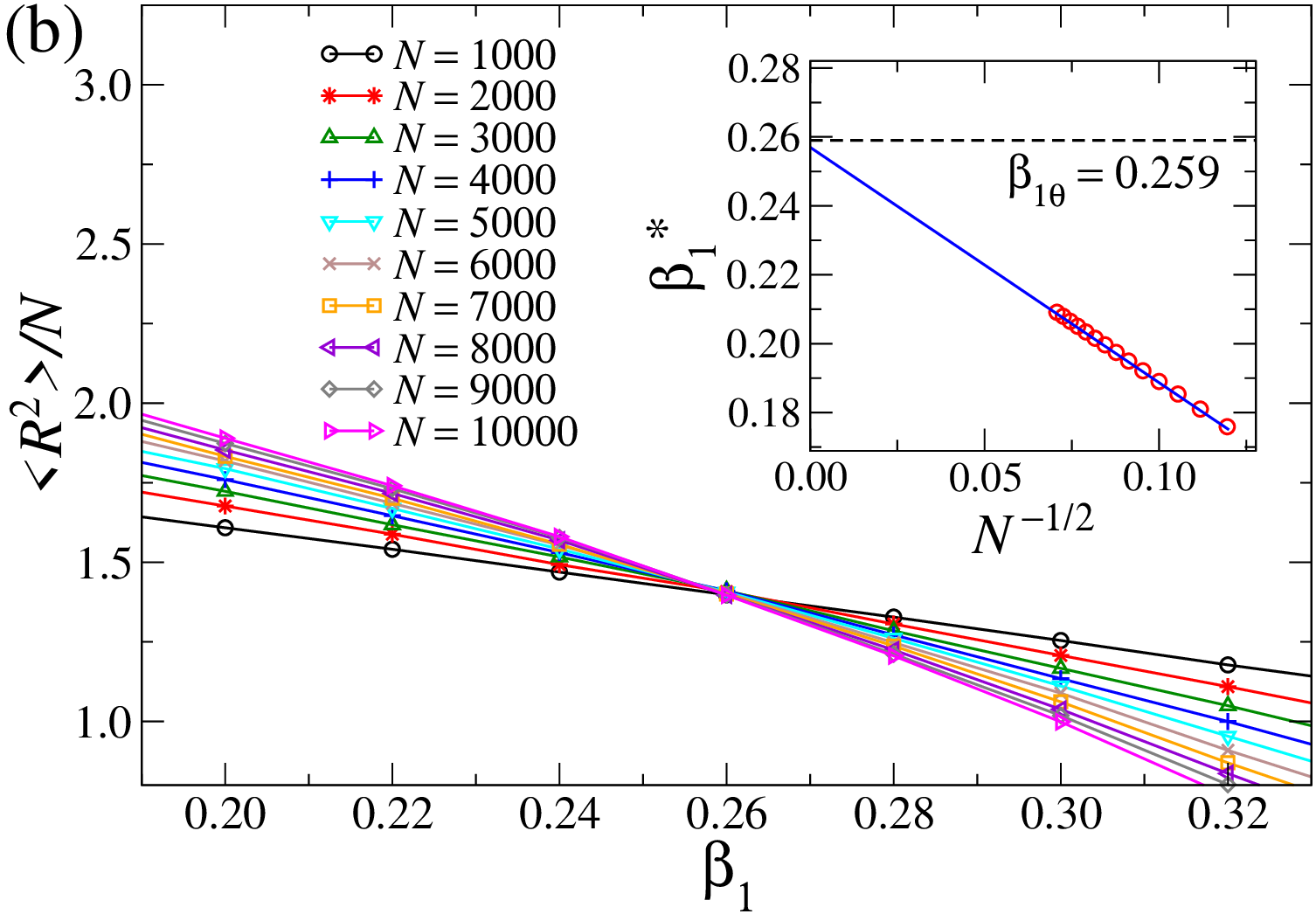}
\caption{(Color online) Rescaled average squared end-to-end distance $\left\langle R^2 \right\rangle /N$ against $\beta_1$ for (a) RA and (b) RF models in 3D with $K=2$. In both panels, the insertions show the extrapolation of $\beta_1^*$ to $N \rightarrow \infty$, where the solid lines are linear fits of the data for the largest chain lengths and the dashed lines show the central values of $\beta_{1,\Theta}$ estimated from the intersections in the main plots.}
\label{fig1}
\end{figure}

First, we analyze the simplest MMS models where each lattice site can be visited by at most $2$ monomers, so that each lattice edge can have at most $3$ and $2$ bonds in RA and RF models, respectively. Figures \ref{fig1}a and \ref{fig1}b show curves of $\left\langle R_N^2 \right\rangle /N^{2 \nu_t} \times \beta_1$ for different chain lengths, with $\nu_t=1/2$, for both models. As already noticed, these curves are expected to intersect each other at the $\Theta$-point, if it exists. This is indeed the case in Fig. \ref{fig1}, where the intersections occur in a very narrow region of $\beta_1$, from which we estimate the $\Theta$-points $\beta_{1,\Theta} = 0.960(15)$ for RA and $\beta_{1,\Theta} = 0.259(9)$ for RF model. (The numbers into parenthesis are the uncertainties.) Very similar values for the $\Theta$-points are obtained from the intersections of curves of $\nu$ exponents against $\beta_1$ for different $N$ (not shown). In this case, beyond $\beta_{1,\Theta}$, the intersections give also estimates of $\nu_t$, being $\nu_t = 0.51(1)$ for both RA and RF models, in striking agreement with the $\Theta$ class.

A final confirmation of the existence of the $\Theta$-points is obtained from the second virial coefficient $A_2$, whose curves (not shown) against $\beta_1$, for a given $N$, cross the zero at a parameter $\beta_{1}^{*}(N)$. Extrapolations of $\beta_1^*(N)$ for $N \rightarrow \infty$ are displayed in the insertions of Figs. \ref{fig1}a-b. The linear behaviors observed confirm the reliability of scaling relation (\ref{eqScalingA2}) and allowed us to estimate $\beta_{1,\Theta}=0.955(8)$ for RA and $\beta_{1,\Theta}=0.257(1)$ for RF model. It is noteworthy that such values, very close to those found from the intersections of $\left\langle R^2 \right\rangle /N^{2 \nu_t} \times \beta_1$ and $\nu \times \beta_1$ curves, were obtained from extrapolations of data for chains of lengths up $N=200$. This indicates that, for these models, possible finite-size corrections to scaling relation (\ref{eqScalingA2}) are negligible. Namely, by assuming $\beta_i^*(N) - \beta_{i,\Theta} = b N^{-1/2} [1+a_1 N^{-\varphi_1}+a_2 N^{-\varphi_2} + \cdots]$, one has $a_i N^{-\varphi_i}\approx 0$ even for small $L$.

At first thought, it seems that immediate reversals of the walks could facilitate the formation of globular configurations and, then, the collapse transition would happen at a smaller parameter $\beta_{1,\Theta}$ in RA model than in RF one. However, the tricritical points found [$\beta_{1,\Theta}^{(RA)} \sim 4 \beta_{1,\Theta}^{(RF)}$] demonstrate an opposite behavior. In fact, as already noticed in Ref. \cite{Tiago}, in RA model walks with a large number of double visited sites are not necessarily collapsed, since they can be formed mostly by sites (say, $j$) visited twice in a sequence $j-k-j$, with $k$ being a NN of $j$. Such sequences, which are obviously absent in RF walks, hinder the formation of globular configurations, due to the larger number of double visited sites created along the RA walks. Namely, these sites give rise to a large effective self-avoidance in RA model and, consequently, a large $\beta_{1,\Theta}$.

\subsection{$K=3$}

\begin{figure}[t]
\includegraphics[width=8.cm]{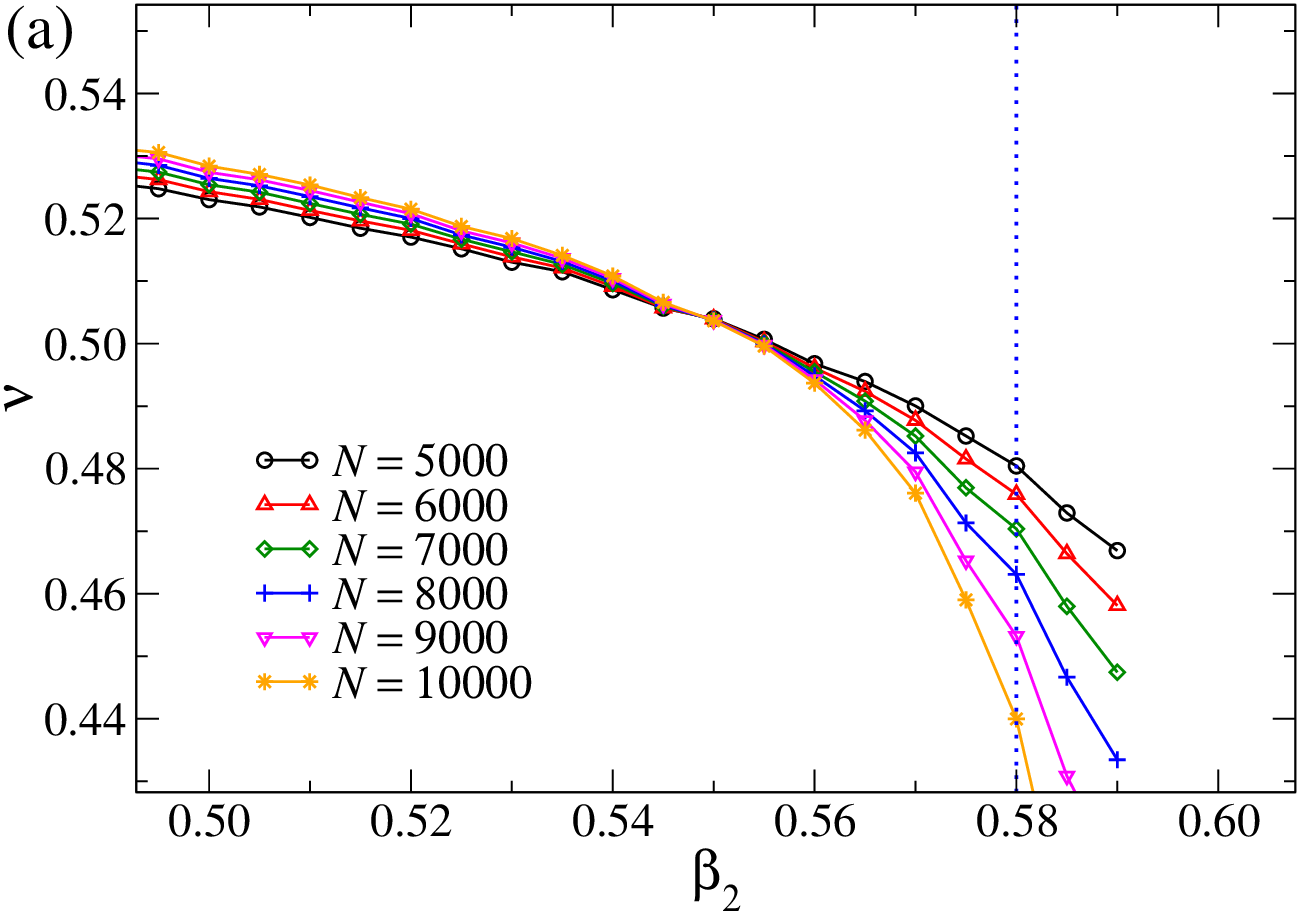}
\includegraphics[width=8.cm]{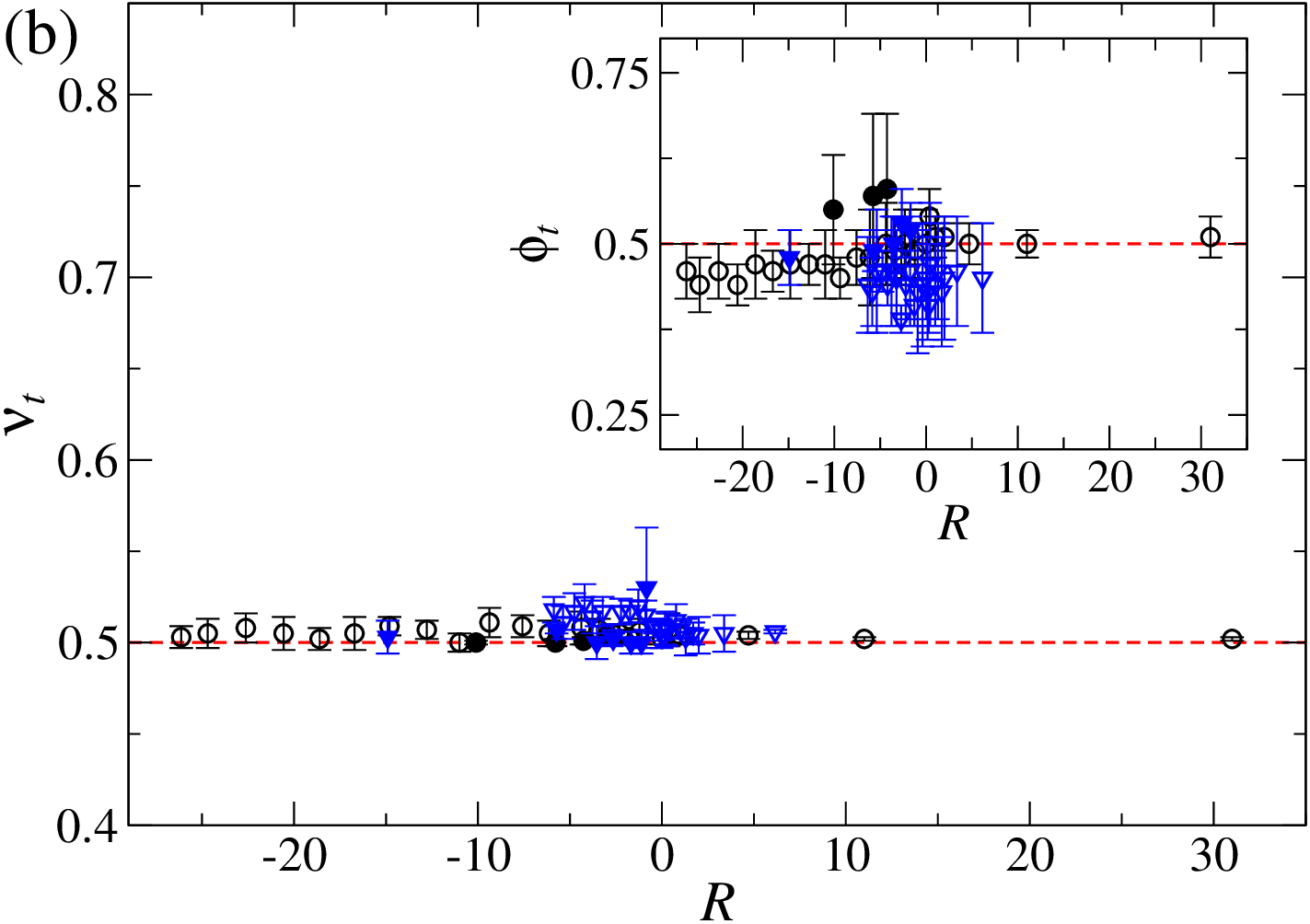}
\caption{(Color online) (a) Exponent $\nu$ against $\beta_2$ for fixed $\beta_1=-0.2$ in RF model in 3D. The vertical (dotted blue) line indicates the value of $\beta_2$ where the probability distribution of sites with three monomers seems to have two peaks (see Fig. \ref{fig5}). (b) Metric $\nu_t$ (main plot) and crossover $\phi_t$ (inset) exponents versus the ratio $R = \beta_{2,\Theta}/\beta_{1,\Theta}$, for RA (blue triangles) and RF (black circles) models in 3D, calculated along the $\Theta$-lines (see Fig. \ref{fig4}). The horizontal (dashed, red) lines indicate the $\Theta$ class exponents, and open (full) symbols are data for $\beta_{1,\Theta} >0$ and ($\beta_{1,\Theta} <0$).}
\label{fig2}
\end{figure}

Now, we investigate the more general models where each site can be visited by at most $3$ monomers, so that each lattice edge can have at most 5 bonds in RA and 3 bonds in RF model. Once, now, one has to deal with two thermodynamic parameters ($\beta_1$ and $\beta_2$), the critical properties are determined (following the same lines from the previous subsection) by keeping $\beta_j$ fixed and varying $\beta_i$, with $\left\lbrace i,j\right\rbrace=\left\lbrace 1,2\right\rbrace$ or $\left\lbrace 2,1\right\rbrace$. For the entire set of parameters analyzed, continuous coil-globule transitions consistent with the $\Theta$ class were found. Namely, curves of $\left\langle R_N^2 \right\rangle /N^{2 \nu_t} \times \beta_i$ for several chain lengths $N$ intersect each other at approximately the same $\beta_i$ (similarly to Figs. \ref{fig1}a-b), when one sets $\nu_t=1/2$. Once again, the coordinates of the $\Theta$-points ($\beta_{1,\Theta},\beta_{2,\Theta}$) obtained from these intersections agree, within the error bars, with those from $\nu \times \beta_i$ curves. An example of such curves is shown in Fig. \ref{fig2}a, for the RF model with fixed $\beta_1 = -0.20$. Similar data are obtained for other values of fixed $\beta_j$ (with $j=1$ or $2$) for both RA and RF models around the transition points. From the intersections of these curves the tricritical metric exponent $\nu_t$ is estimated, whose values are depicted in Fig. \ref{fig2}b as function of the ratio $R \equiv \beta_{2,\Theta}/\beta_{1,\Theta}$. The excellent agreement of these exponents with the mean-field value ($\nu_t=1/2$) strongly suggests that the coil and globule phases are separated by lines of $\Theta$-points ($\Theta$-lines), in both RA and RF models, in the broad range of parameters analyzed.

\begin{figure}[t]
\includegraphics[width=8.cm]{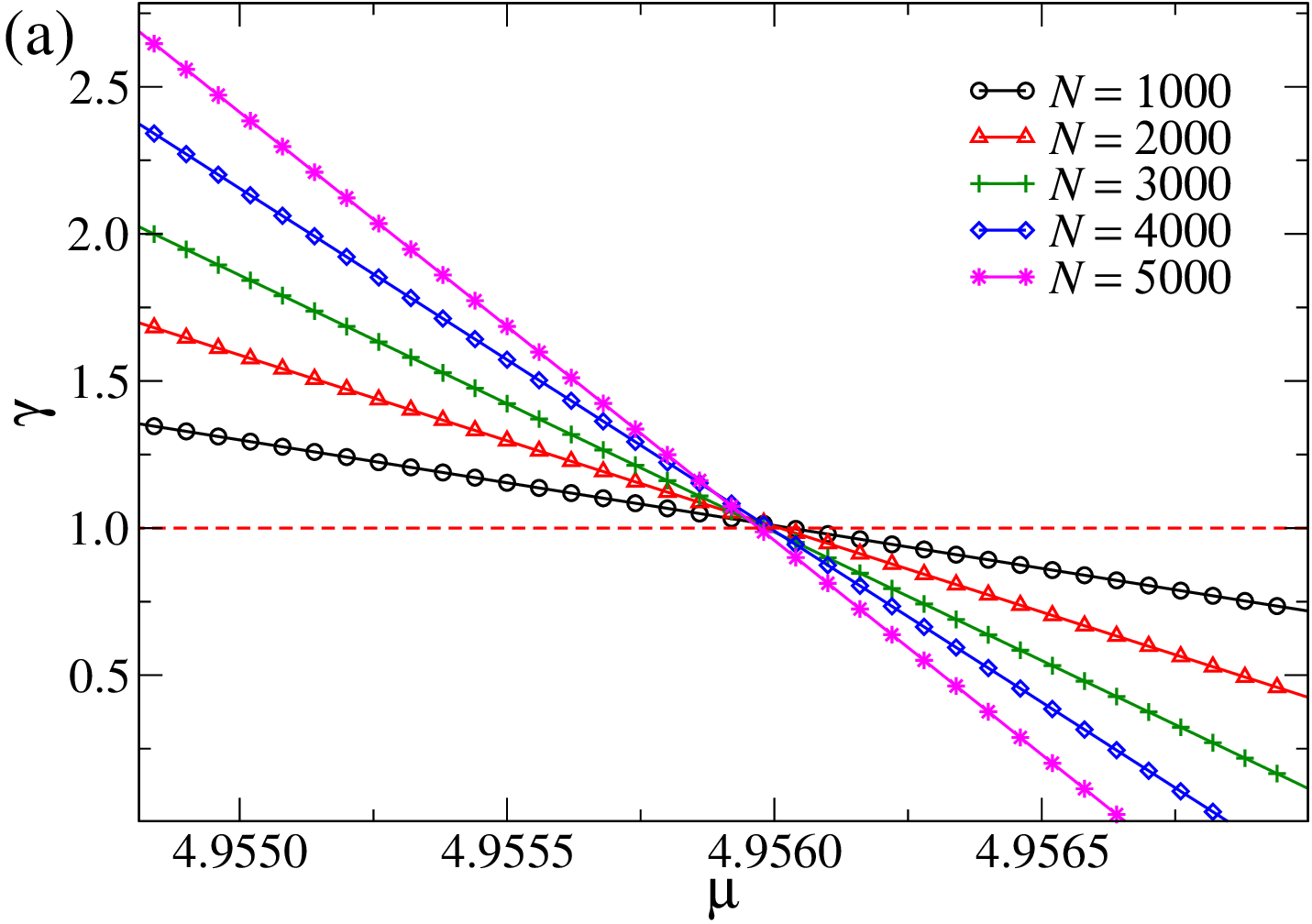}
\includegraphics[width=8.cm]{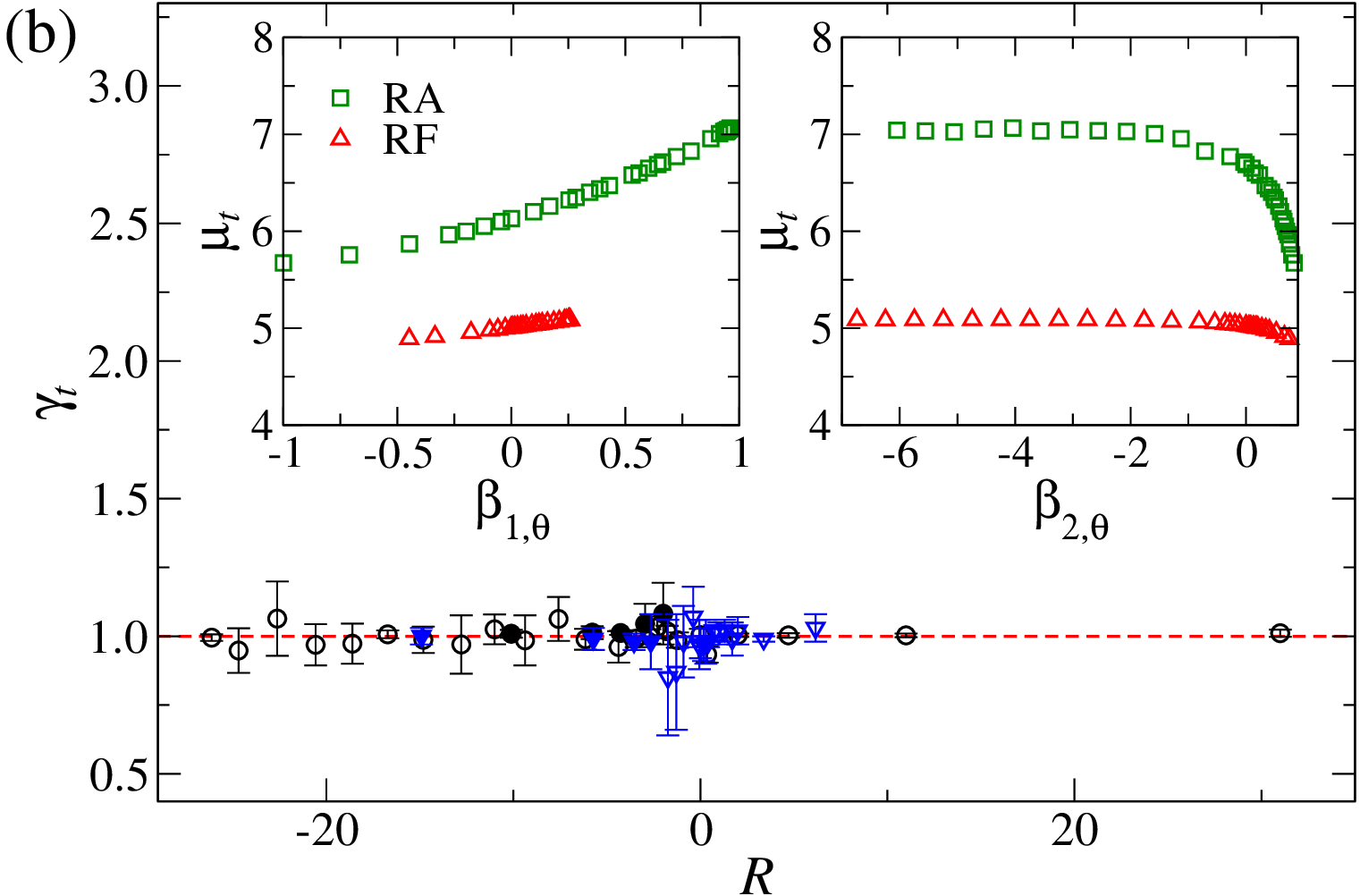}
\caption{(Color online) (a) Exponent $\gamma$ versus the connectivity coefficient $\mu$ calculated at the $\Theta$-point $(\beta_{1,\Theta},\beta_{2,\Theta})=(-0.2,0.55)$ for the RF model in 3D. (b) Tricritical entropic exponent $\gamma_t$ against the ratio $R = \beta_{2,\Theta}/\beta_{1,\Theta}$ for RA (blue triangles) and RF (black circles) models in 3D. In both panels, the horizontal (dashed red) line indicates the $\Theta$ exponent $\gamma_t=1$. In (b), open (full) symbols are data for $\beta_{1,\Theta} >0$ and ($\beta_{1,\Theta} <0$), and the insertions display the connectivity constants $\mu_t$ as functions of $\beta_{1,\Theta}$ (left) and $\beta_{2,\Theta}$ (right) for both RA and RF models in 3D.}
\label{fig3}
\end{figure}

Along the transition lines, the crossover exponents $\phi_t$ are estimated from derivatives of $\ln\left\langle R_N^2\right\rangle$, following Eq. \ref{eqScalingphi}. Such exponents are displayed in the insertion of Fig. \ref{fig2}b as a function of the ratio $R$. In contrast with $\nu_t$, the central values of $\phi_t$ agree well with the $\Theta$ value ($\phi_t=1/2$) only for $R>0$, presenting larger fluctuations for $R<0$. One possible explanation for this are the larger uncertainties in the loci of the $\Theta$-points in the regions of repulsive interactions (see Fig. \ref{fig4} below). Moreover, since we are neglecting logarithmic corrections when estimating the exponents, some small deviations can be expected. Anyhow, the $\phi_t$ exponents are always close to $1/2$ and far from $1$, discarding the possibility of discontinuous coil-globule transitions in MMS models.

At each tricritical point, curves (for different $N$) of the entropic exponents $\gamma$ as function of the connectivity coefficient $\mu$ are expected to intersect at a single point, from which the tricritical exponent $\gamma_t$ is estimated, as well as the value of the connectivity $\mu_t$. Figure \ref{fig3}a shows an example of this kind of plot, where indeed one observes the expected behavior (similar plots are found along the entire transition lines). The values of $\gamma_t$ obtained from these intersections are depicted in Fig. \ref{fig3}b against the ratio $R$, where the variation of $\mu_t$ with $\beta_{1,\Theta}$ and $\beta_{2,\Theta}$ is also shown. For both RA and RF models, the exponents agree quite well with the mean-field value $\gamma_t=1$, providing additional confirmation of the $\Theta$ universality class of the transitions. 

The values of $\mu$ at the coil-globule transition are larger in RA model than in RF one, as expected. Interestingly, while $\mu_t$ is always smaller than the lattice coordination ($q=6$) in RF model, for the RA it becomes larger than $q$, and saturates at $\mu_t \approx 7$ for large $\beta_{2,\Theta} < 0$. We note that $\mu_t > q$ have also been observed in a generalized ISAW model with competing NN and next-NN interactions \cite{Nathann14}. For large negative $\beta_{2,\Theta}$, the saturation value for the RF model is $\mu_t \approx 5.08$, which is intriguingly close to the $\Theta$ point value for the ISAW model: $\mu_t^{(ISAW)} \approx 5.04$ \cite{Nathann14,meirovitch2}.

Figures \ref{fig4}a and \ref{fig4}b display the phase diagrams for the RA and RF models, respectively. In both models, the $\Theta$-lines for negative $\beta_2$ converge to asymptotic values corresponding to those of the $K=2$ models, what is expected since the MMS models with $K=2$ correspond to the limit $\beta_2 \rightarrow -\infty$ of the $K=3$ case. This strongly suggests that both $\Theta$-lines shall extend to $\beta_{2,\Theta} \rightarrow -\infty$.

\begin{figure}[t]
\includegraphics[width=8.cm]{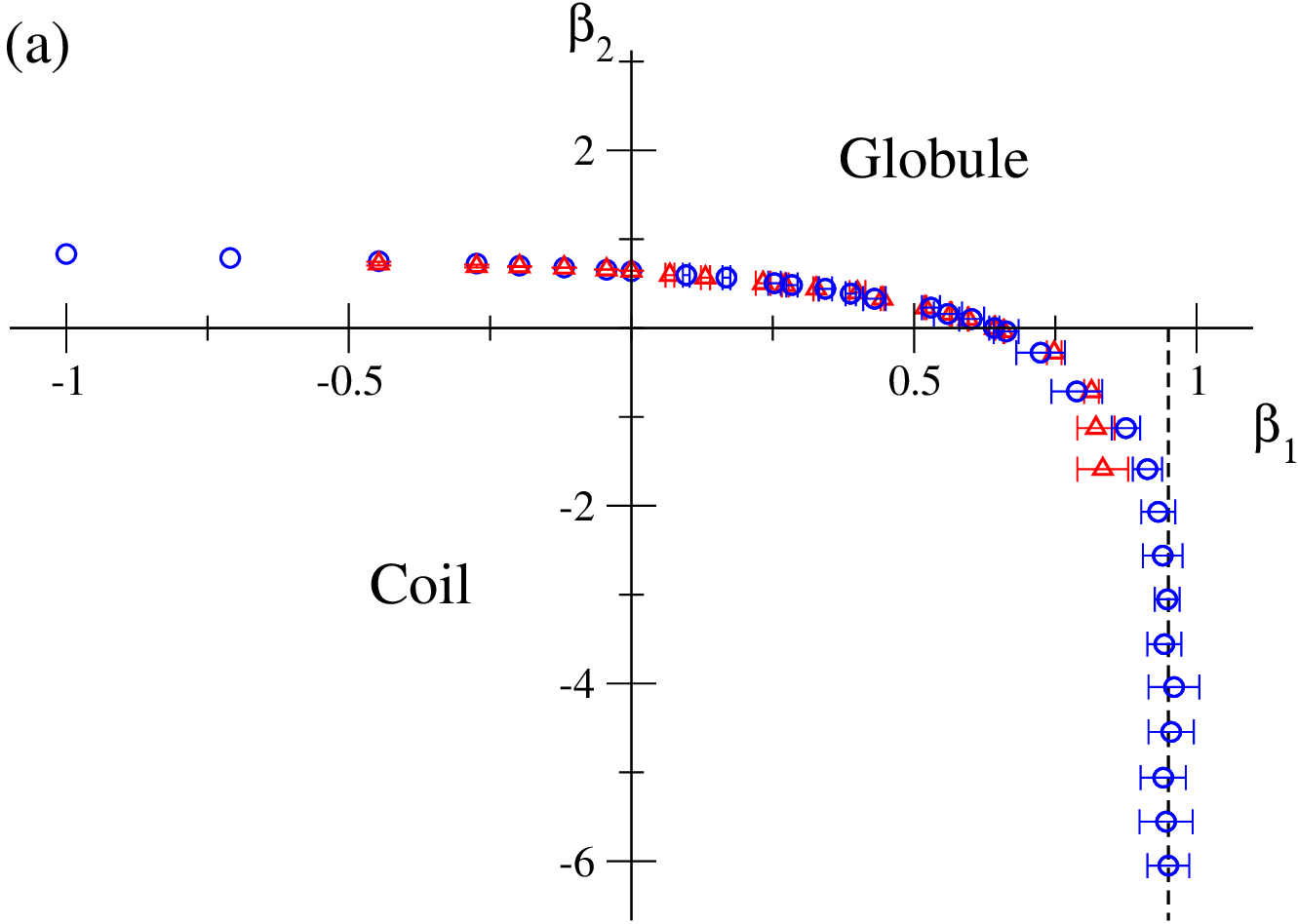}
\includegraphics[width=8.cm]{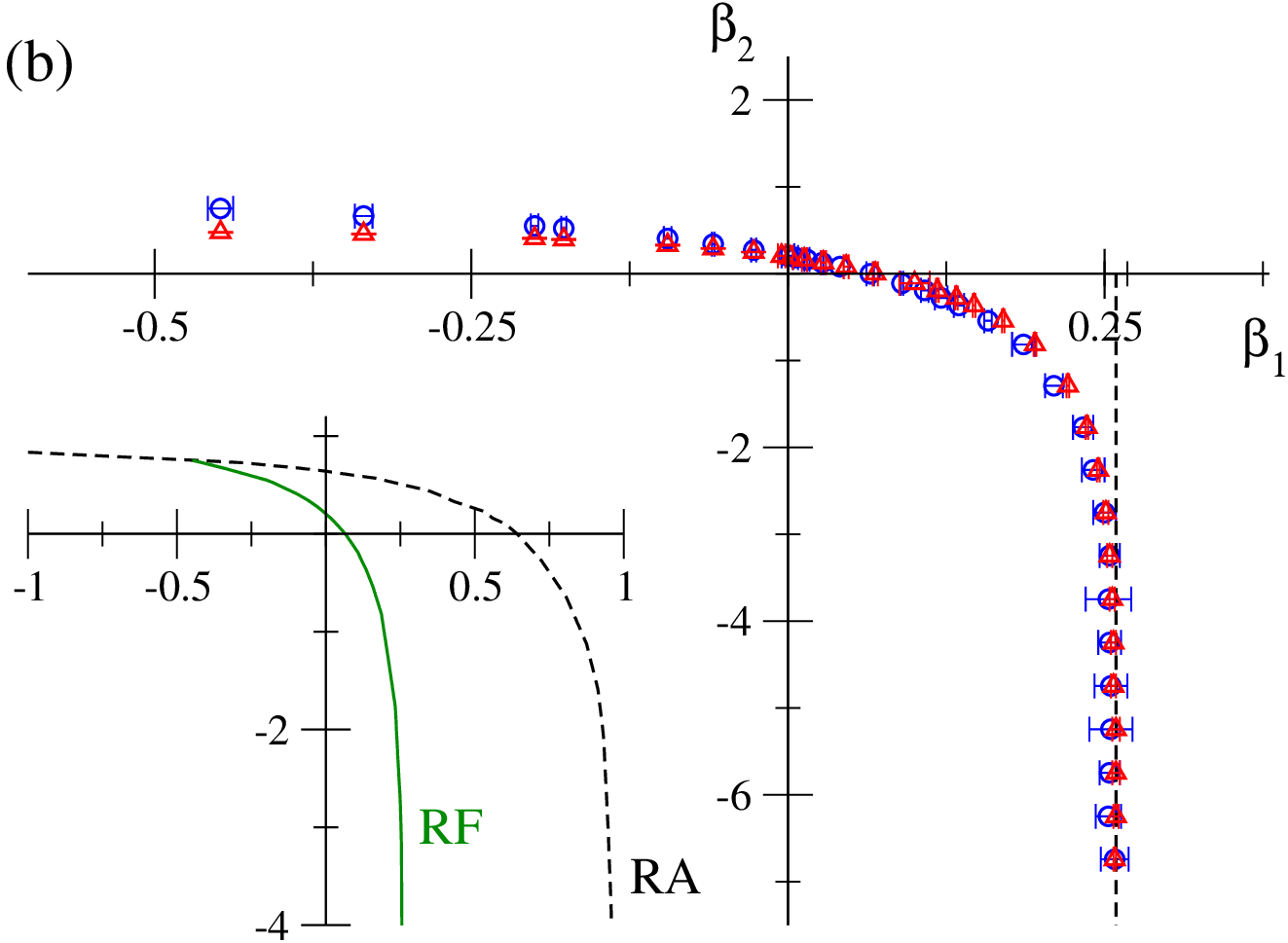}
\caption{(Color online) Phase diagrams for (a) RA and (b) RF models in 3D with $K=3$. In both panels, circles (blue) and triangles (red) are the tricritical points obtained respectively from the intersections of $\left\langle R_N^2 \right\rangle/N$ curves and from the second virial coefficient. The dashed vertical lines (black) indicate the values of $\beta_{1,\Theta}$ for the case $K=2$. A comparison between the $\Theta$-lines for RA and RF models is shown in the insertion in (b).}
\label{fig4}
\end{figure}

The $\Theta$-lines seem also approach asymptotic limits for large negative $\beta_1$. However, with PERM and other standard Monte Carlo methods to generate polymer chains, it is not possible to demonstrate this by sampling the limiting case $\beta_1 \rightarrow -\infty$, where sites with two monomers (dimers) are forbidden while sites with three monomers (trimers) are allowed. The reason is obviously that to achieve a conformation with trimers the walk should first pass through (forbidden) configurations with dimers. Indeed, we observe that the more negative $\beta_{1}$ becomes the large the fluctuations in the data are and more samples are need to yield reliable results for the transition. This problem, which is worse in RF model, has limited our analysis to $\beta_1 \gtrsim -1$ in RA and $\beta_1 \gtrsim -1/2$ in RF model. This is certainly related to the difficult of sampling characteristic conformations of large $\beta_1 < 0$ (and $\beta_2 > 0$) - i.e, with high (low) density of trimers (dimers) - with chain growing methods.

The insertion in Fig. \ref{fig4}b shows a comparison of the $\Theta$-lines for RA and RF models. For $\beta_{1,\Theta} \gtrsim -0.45$ the coordinates of the $\Theta$-points for the RA model are always larger than those for the RF one, showing that the coil-globule transition is facilitated when immediate reversals of the walks are forbidden, as already discussed in the previous subsection. For $\beta_{1,\Theta} \lesssim -0.45$, notwithstanding, an inverse situation seems to arise, with $\beta_{2,\Theta}$ becoming larger for the RF model. Though with our data we can only infer this, one remarks that such scenario have indeed been found in the mean-field solutions of these models on the Bethe lattice \cite{Tiago}. For instance, for a Bethe lattice with coordination $q=6$, the tricritical line in the limit $\beta_{2} \rightarrow -\infty$ for RA (RF) model is located at $\beta_{1}\approx 0.53$ ($\beta_{1} = 0$), whilst the line of critical-end-points in the limit of $\beta_{1} \rightarrow -\infty$ is at $\beta_{2} \approx 0.86$ ($\beta_{2} \approx 1.13$) \cite{Tiago}. In fact, for large $\beta_1 <0$, the creation of dimers is quite difficult and, since the immediate reversals of the walks contribute to this, the collapse transition turns out to be facilitated in RA model.

Figures \ref{fig4}a-b also show the $\Theta$-lines obtained from extrapolation of the zero points of the second virial coefficient $A_2$. In general, the agreement between these curves and the ones obtained from the intersections in rescaled data is quite good, giving a final and undoubted confirmation that the coil-globule transitions in the MMS models are continuous and belong to the $\Theta$ class. One notices that in RA model strong fluctuations arises in $A_2$ for large $\beta_2 < 0$, preventing a reliable estimate of the $\Theta$-points, whereas RF model presents a smooth behavior in such region. In opposition, for large $\beta_1 < 0$, one observes that corrections to scaling (\ref{eqScalingA2}) are more severe in RF model. Indeed, in this region the $\Theta$ points estimated from $A_2$ are systematically smaller than those from the intersections in RF case. These problems are possibly due to the small walks' lengths considered to calculate $A_2$ ($N=200$), as opposed to $N=10000$ used in the other analyses. Unfortunately, the numerical procedure to obtain $A_2$ is very demanding computationally and thus investigating this quantity for long walks is very hard.

\begin{figure}[t]
\includegraphics[width=8.cm]{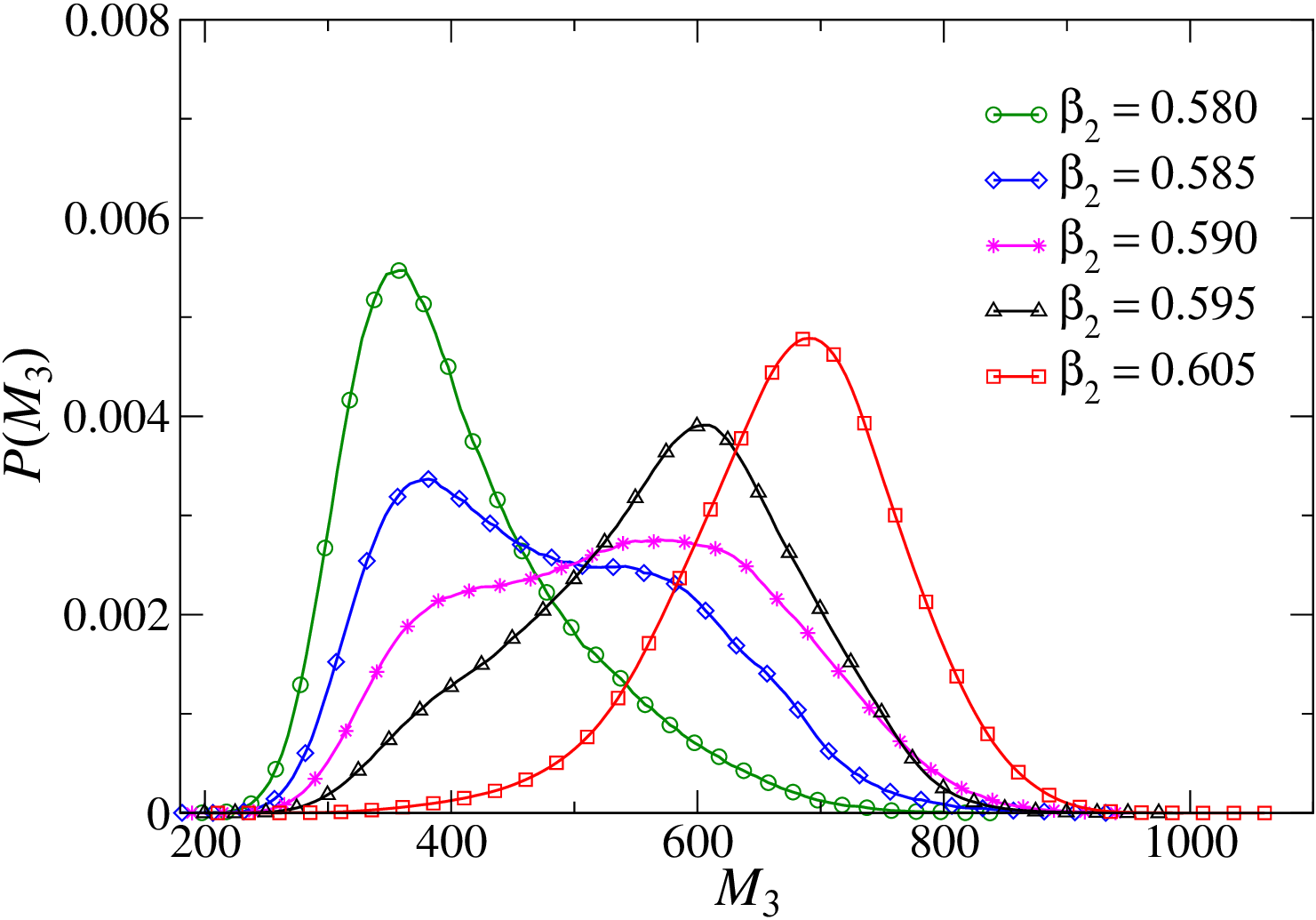}
\caption{Probability distributions of the number of sites visited by three monomers, for the RF model in 3D with fixed $\beta_1=-0.2$ and several values of $\beta_2$.}
\label{fig5}
\end{figure}

At this point, we remark that these problems with $A_2$ are, in some sense, consistent with previous MC results for the MMS models \cite{Krawczyk}. For instance, for the RA model in the region of negative $\beta_2$, Krawczyk \textit{et al.} \cite{Krawczyk} were not able to determine whether a coil-globule transition exists or not. Moreover, for $\beta_1 < 0$, they suggest that the transition in RF model is discontinuous, which could be explained by the strong finite-size corrections found here (in the \textit{continuous} transition). Actually, the conclusion that a fist-order transition exists for $\beta_1 < 0$ was also based on probability distributions with a ``weak bimodality'' in \cite{Krawczyk}. Indeed, we found evidence of a \textit{very weak} bimodality in the distributions for the number of sites with trimers [$P(M_3)$] (as well as with dimers and monomers) in RF model with $\beta_1<0$. An example of this is shown in Fig. \ref{fig5} for fixed $\beta_1 = -0.2$, where only pronounced ``shoulders'' are seem in the distributions, rather than two well-defined peaks. It is noteworthy that such ``shoulders'' appear only in a very tiny range of parameters, being for example $0.585 \lesssim \beta_2 \lesssim 0.595$ in Fig. \ref{fig5} (see also Fig. 4 in Ref. \cite{Krawczyk}). Moreover, we do not find any evidence of a building up of bimodality as the polymer length increases, indicating that these ``shoulders''/peaks are not related to a true phase coexistence. We notice that if a coexistence line would exist here, it would be very close to the $\Theta$-line. For instance, for $\beta_1=-0.2$ one has $\beta_{2,\Theta} = 0.550(3)$ and the coexistence would be at $\beta_{2,c} \approx 0.58$ (see Fig. \ref{fig2}a), and the same would happen for other parameters. We remark that a coil-globule transition very close to a first-order ``globule-crystal'' transition have been reported for the bond fluctuation model for finite flexible chains \cite{Rampf1,Rampf2,Paul1,Paul2}. However, the rising of a stable ordered (crystalline) phase in the MMS models, without the addition of any local chain stiffness on them, seems quite unexpected. In fact, no evidence of the existence of a third phase in the canonical phase diagrams was found here or elsewhere \cite{Krawczyk,Pablo,Tiago}.

\section{Results for the square lattice}
\label{results2D}

Now, we turn to the analysis of the MMS models on the square lattice. 

\begin{figure}[t]
\includegraphics[width=8.cm]{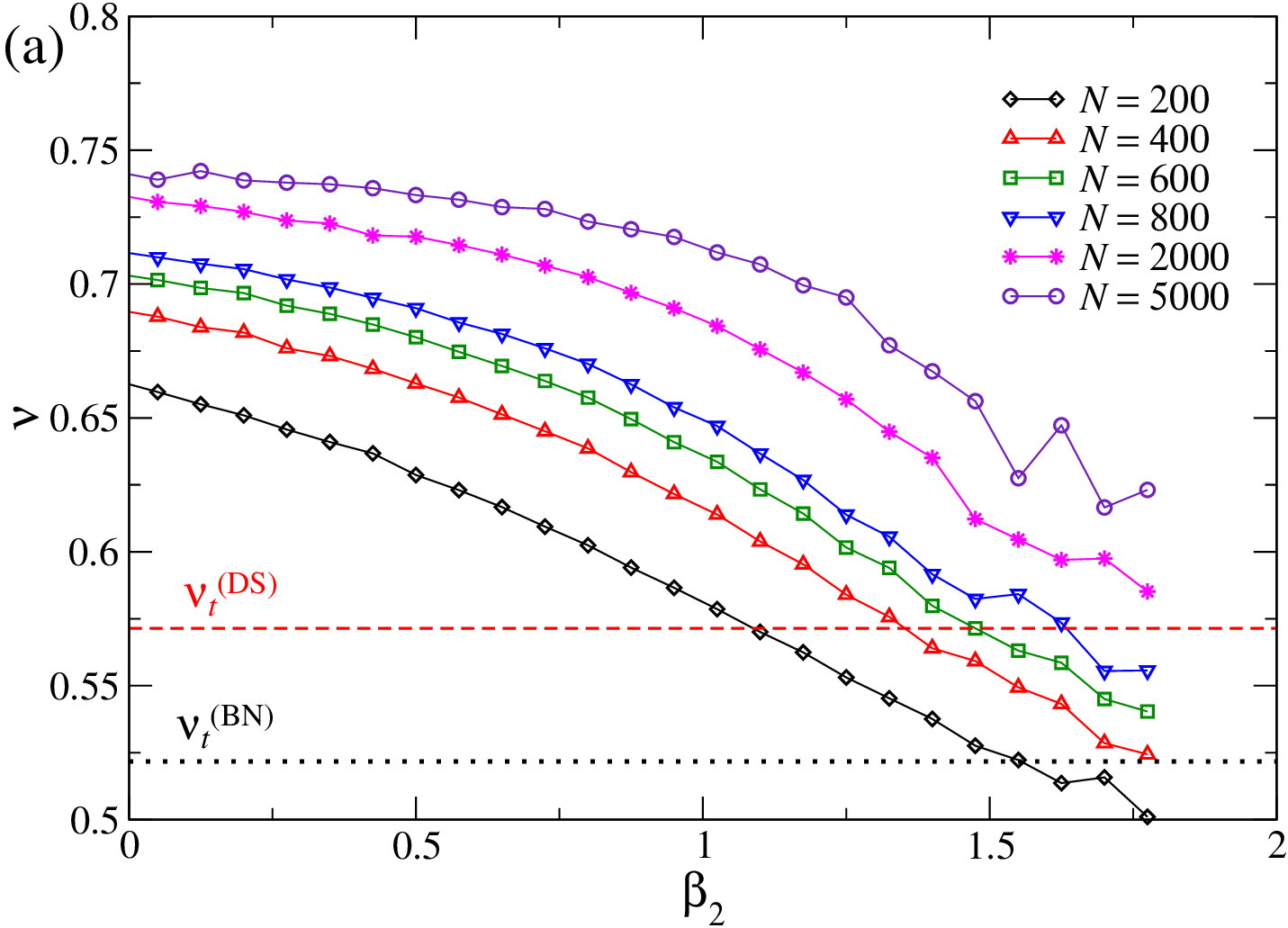}
\includegraphics[width=8.cm]{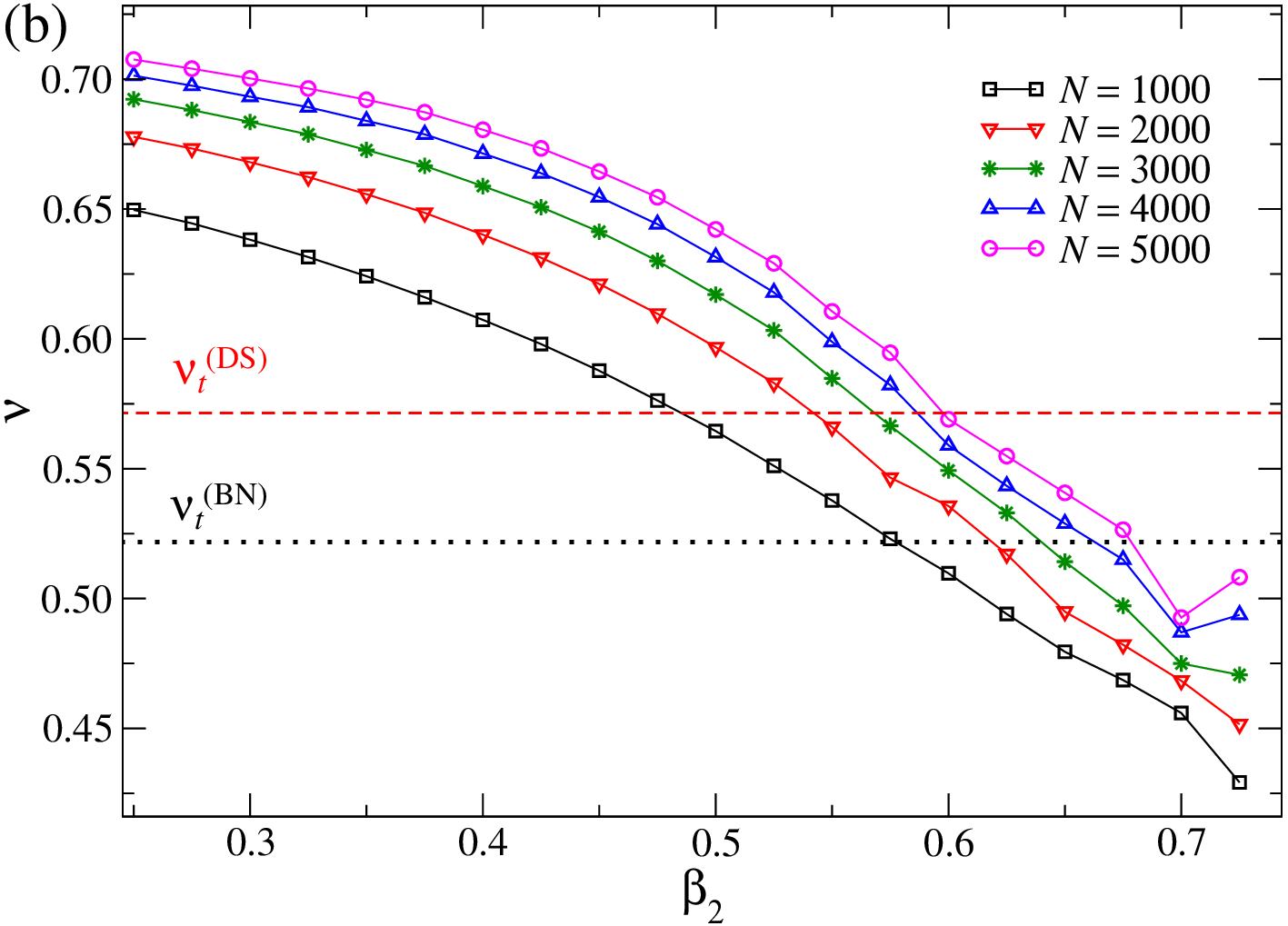}
\caption{Metric exponents $\nu$ against $\beta_2$ for (a) RA and (b) RF models in 2D and several chain lengths. The parameter $\beta_1$ is fixed at $\beta_1=0$ in (a) and $\beta_1=-0.20$ in (b). The dashed and dotted horizontal lines indicate the exponents expected for the Duplantier-Saleur (DS) and Bl\"ote-Nienhuis (BN) universality classes.}
\label{fig6}
\end{figure}

Once again, the same thermodynamic behavior is observed in RA and RF models. In contrast to the 3D case, however, here we do not find a coil-globule transition. Instead, a smooth crossover seems to exist between these phases. An evidence of this is obtained from the variation of the $\nu$ exponents with $\beta_i$, for a fixed $\beta_j$, whose curves for different lengths do not intersect at any point. Examples of this behavior are shown in Figs. \ref{fig6}a and \ref{fig6}b for the RA and RF models, respectively, and similar results are found for a broad range of parameters analyzed: $-1 \lesssim \beta_1 \lesssim 1$ and $-1 \lesssim \beta_2 \lesssim 1$, as well as the limit $\beta_2 \rightarrow -\infty$ (the $K=2$ case). The absence of intersections, and consequently of a $\Theta$ behavior, is also observed in plots of $\left\langle R_N^2 \right\rangle/N^{2\nu_t}$ if one assumes that $\nu_t=\nu_{DS}=4/7$ or $\nu_t=\nu_{BN}=12/23$.

This result for the RA model agrees, for the first time in this work, with the findings by Krawczyk \textit{et al.} \cite{Krawczyk}. For the RF model, however, it was claimed in \cite{Krawczyk} that a coil-globule transition exists for $\beta_1 < 0$, while our analysis strongly suggests that it is absent also in this case, at least for $\beta_1 \gtrsim -1$.

\section{Final discussions and conclusions}
\label{conclusions}

We have presented an extensive numerical analysis of the polymer models by Krawczyk \textit{et al.} \cite{Krawczyk}, where lattice sites can be visited by up to $K=3$ monomers and Boltzmann weights $\omega_0=1$, $\omega_1=e^{\beta_1}$ and $\omega_2=e^{\beta_2}$ are associated, respectively, to sites occupied by $1$, $2$ and $3$ monomers. In these multiple monomer per site (MMS) models the maximal number of bonds per lattice edge ($N_b$) can be larger than one and depends on whether immediate reversals of the walks are allowed (RA, where $N_b = 2K-1$) or forbidden (RF model, where $N_b=K$). This is a key distinction of these systems from other classical lattice models used to investigate the polymer collapse transition, such as ISAW, ISAT and VISAW, where $N_b=1$. Obviously, systems with $N_b>1$ have an ensemble of walks much closer to simple Random Walks than those with $N_b=1$. Moreover, $N_b>1$ turns the lattice system highly non-planar in 2D, what could explain the absence of a coil-globule transition in the MMS models on the square lattice, as observed here for both RA and RF models and also previously for the RA one \cite{Krawczyk}. However, the effect of the lattice coordination $q$ can also be playing an important role at this point, so that further numerical investigations of these models on other lattices (e.g., the triangular one) are worthy. Moreover, we believe that these results shall motivate the development of field theories mapping on walks on 2D lattices with multiple bonds per edge, in order to verify whether this can yield a breakdown of the collapse transition.

In the simple cubic lattice, we have found strong evidences that coil and globule phases are always separated by a $\Theta$-line, in both RA and RF models, as confirmed by several scaling exponents and extrapolations of the zero's ($\beta_i^*$) from the second virial coefficient ($A_2$). This scenario is different from that suggested in previous simulations of these models \cite{Krawczyk}, with continuous (discontinuous) transitions in the region of $\beta_2 < 0$ ($\beta_1 < 0$) in RF model, while in the RA one for $\beta_2 < 0$ those authors were not able to decide between a transition or a simple crossover. Indeed, our data for $A_2$ (for $N\leqslant 200$) deep inside the region of $\beta_2<0$ in RA model present strong fluctuations, which prevent us from performing reliable extrapolations of $\beta_i^*$. This suggests that for this model/region short walks do not present the equilibrium conformations expected for this parameter set. This seems to be confirmed by the fact that for long walks our results (for other quantities) in the same model/region are well-behaved. In the RF model with $\beta_1<0$, rather than large fluctuations in $\beta_i^*$, strong corrections to scaling relation (\ref{eqScalingA2}) are found, in a way that large effective crossover exponents are observed for very short chains. This may explain the strong buildup of fluctuations (in the number of sites with trimers) reported in Ref. \cite{Krawczyk}. Namely, all these results point out that the difficult in determining the existence of the transitions, as well as the first-order one claimed in Ref. \cite{Krawczyk} can be consequences of the not so large chain lengths investigated there.

Finally, we notice that in some respect the $\Theta$-lines in RA and RF models in 3D are consistent with the behavior of these models on the Bethe lattice. Even though for $\beta_1<0$ the tricritical lines give place to critical-end-point (CEP) lines in such lattice, it is pretty possible that in such mean-field approaches the order of the coil-globule transition is being underestimated (from tricritical to CEP). To further confirm this, it would be worthy developing methods for investigating the interesting limit $\beta_1 \rightarrow -\infty$, where dimers are forbidden and trimers are allowed in the walks, since with known methods it seems not possible to study this case.


\acknowledgments

The authors thank T. Prellberg for helpful discussions, A. L. Owczarek for a critical reading of the manuscript and the support from CNPq, Capes and FAPEMIG (brazilian agencies).

\end{document}